\documentclass[aps,prl,twocolumn,groupedaddress]{revtex4}
\usepackage{graphicx}

\begin{document}
\title{Imaging the charge transport in arrays of CdSe nanocrystals}
\author{M.~Drndic$^1$, R.~Markov$^1$, M.V.~Jarosz$^2$,
M.G.~Bawendi$^2$, M.A.~Kastner$^1$,N.~Markovic$^3$, M.Tinkham$^3$}
\affiliation{$^1$Department of Physics, $^2$Department of Chemistry, MIT,
Cambridge, MA}
\affiliation{$^3$Department of Physics, Harvard University, Cambridge, MA}

\date{\today}

\begin{abstract}
A novel method to image charge is used to measure the diffusion coefficient
of electrons in films of CdSe nanocrystals at room
temperature. This method makes possible the study of charge transport in
films exhibiting extremely high resistances or very small diffusion
coefficients.
\end{abstract}

\pacs{}

\maketitle
Electric-force microscopy (EFM) \cite{Bonnell} has been used to probe the electrical 
properties of microscopic systems, such as DNA molecules
\cite{Bockrath}, carbon nanotubes \cite{Bachtold}, SiGe dots \cite{Tevaarwerk}, and single 
nanocrystals (NCs) \cite{Krauss}.  However, there are few examples in
which EFM data are analyzed quantitatively \cite{Krauss}. 

We describe a new method, based on EFM \cite{Bonnell}, to image the charge
motion in a regime not
accessible to traditional transport measurements.  We use it to
directly obtain the
charge diffusion coefficient in films
of CdSe NCs.
Semiconducting NC arrays made to date are
so resistive that it is difficult to study them.
The current through such arrays after application of a voltage step
decays with time \cite{Morgan,ginger}, so that the steady-state current corresponds to a resistance that is too large to
measure by conventional means.
Our technique could have wide application in
highly resistive thin films.

The charge transport is imaged in three-dimensional arrays of CdSe
NCs using a field-effect transistor geometry \cite{Morgan}.  The EFM
measurements are made after the voltage on the source electrode is
turned on and off.  The device, illustrated in Fig. 1. 
consists of two Au electrodes separated by $\sim 5 \mu$m, a
350nm thick $SiO_{2}$ layer, and the degenerately doped Si substrate.
Highly monodispersed NCs, capped with tri-octylphosphine oxide with $\approx$ 1.1 nm between
nearest neighbors \cite{Murray}, are self-assembled into $\sim$ 0.2 $\mu$m thick 
films on top of the devices \cite{Morgan}. We show data for two 
samples, with NC diameters 6.1 nm (Fig. 2) and 4.7 nm (Fig. 4). 
These NCs are synthesized using different preparations, 
as described in \cite{Murray} and \cite{Nathan}, respectively.
Negative dc-voltage $V_{dc}$ is applied to
the source electrode to charge the NCs, while all other voltages are 
zero (Fig.1).
Our approach is different from the standard EFM technique in which an 
ac-voltage is applied to the tip \cite{Bonnell}.

We first discuss how the EFM provides a measure of the charge
distribution. A small conductive
tip oscillates in the z-direction and is scanned parallel
to the sample surface (Fig.1) . The electrostatic force $F$ between the sample
and the tip modifies its resonant frequency and introduces a
phase shift $\Delta{\phi}(x,y) \propto {dF/dz}$ 
\cite{Bonnell}. An image is
produced by measuring the phase shift
as a function of the tip position in the
xy-plane at a constant height.
The electrostatic force on the tip results
from the capacitive coupling between the tip and the electrodes (or
the gate) and from the Coulomb interaction between the static
charges on the sample and the charges induced on the tip and the electrodes
(or the gate). Krauss {\it et al.}~\cite{Krauss} have discussed the force
from a point charge.  We use a similar argument before extending it to 
a charge distribution in the film.  
The charge $q$ induces $q_{1}$ on the gate and $q_{2}$ on the tip. 
Following \cite{Krauss}, $q_{1},q_{2} \propto q$, and ignoring contact potentials
because we work with large voltages, one can show that $F \propto q^{2}$ for $V_{dc}=0$. The EFM signal is then 
$\Delta{\phi} \propto q^{2}$, proportional to the square of the charge.

We argue that the EFM signal for a continuous charge distribution
is similarly proportional to the square of the charge in a small area of
the film below the tip.  We scan the tip $\sim400$ nm above the film-SiO$_2$
interface where the charge density resides. Because the oxide is $\sim350$ nm thick,
the image charge density in the gate is $\sim350$ nm below that in the
film.  Thus, at lateral distances larger than $\sim400$ nm the charge
density in the film will be equal and opposite to the image charge in the
Si substrate, giving a very small net force on the tip.  The tip is
only sensitive to charge within a radius of $\sim400$ nm.  Considering only
charges q, q$_1$ and q$_2$ within this radius, the argument
above applies and the force is once again proportional to $q^2$.

We use a Digital Instruments atomic-force microscope with conducting tips. 
The tip scans twice over each line along x in the xy plane above the
sample. In the first pass, the tip traces the topography by scanning close to the surface \cite{Bonnell}. 
In the second pass above the same line, the tip
is raised to a larger height (200 nm).  Using the topographic profile, the tip is maintained at a constant height while the EFM signal is
recorded. Each scan takes $\approx 3 min$.
At the end of each scan, the tip is moved back to the top of the
image before the next scan. Because of significant spatial drifts over long
time periods, we scan the topography of every
line before each EFM scan. Samples are measured in a nitrogen gas atmosphere to
prevent degradation.  

Figure 2(a) shows the time evolution of EFM images with $V_{dc} =
-40V$ applied to the source electrode. The topographical countour of the 
same region containing the electrodes is superimposed.
$V_{dc}$ is changed from $0V$ to $-40V$ at the top of the
first scan, and it is held constant during the following scans.
The second image is recorded $39$ minutes after the first image.  As expected, the grounded drain electrode
cannot be seen in the EFM image. The EFM signal changes in time and the
charged region around the source grows as the charge spreads into the film (Fig. 2(a)). Similar
measurements on identical devices without depositing a film show
no growth of the charged
region at the electrode, guaranteeing that the charge motion is a property
of the film.

Figure 2(b) shows the time evolution of the EFM images as
the film discharges after the voltage on the source is set back to zero.  Before the first line scan, the film has
been charged for $\approx 60$ minutes at $V_{dc} = -40V$.  Note that the voltage has been set to
zero shortly after the scan has begun.
The second scan is initiated 9 minutes after the first image.
We have imaged samples
charged for up to $\sim 4 hr$, required for the charge
to reach the drain,
and discharged over approximately the same time period.

During charging, the strong electric force near the
source electrode influences the topographic contour making it impossible
to measure the force at constant height.  However, the force during
discharging is much smaller, and we believe that discharging images
directly give the square of the charge
distribution in the film.

Figure 3 shows a series of line scans along the x-axis
for a constant $y$, corresponding to Fig. 2 (b).
The data are symmetrical around the center of the electrodes, and Fig.3
shows the data on one side of the electrode; the edge of
the electrode is set to $x=0$. We show the line scan at $y = 0.3\mu{m}$, 
close to the top of the images in Fig. 2 (b).
Line scans are shown for times $t_{1} = 0.05 min$, and
$t_{j}=t_{1}+3(j-1){(min)}$, $j = 1, 4, \dots,19$.
The voltage on the source is set to zero at $t = 0$. 
The narrow peaks in the EFM signal come from the roughness in the
topography.

The calculated curves $(Q(x,t_{j}))^{2}$ are obtained by fitting a diffusion
model simultaneously to all the measured curves for
$j= 1$ to $21$. We have used the analytical solution of a
one-dimensional (1D) diffusion equation along x, applicable
for line scans corresponding to positions far from the tip of
the electrode. We first calculate the diffusion of charge from the
electrode into the film:
$Q_{ch}(x,t)={Q_{0}}{\left[1-Erf\left({x}/{(2\sqrt{{D}{t}})}\right)
\right]},$
for $x\geq 0$ and  $t\geq 0$, where
$Erf(\alpha)=2/{\sqrt{\pi}} \int_{0}^{\alpha}{e^{-s^{2}}ds}$,
and $D$ is the diffusion coefficient.
This solution satisfies $Q(x,t)=
Q_{0}$ for $x=0$, $t\geq0$, and $lim_{x\rightarrow \infty} Q(x,t) =
0$.
The charge distribution during discharging is
$Q_{disch}(x,t)=Q_{ch}(x,t+\tau_{ch})-Q_{ch}(x,t)$,
where $x\geq 0$, $t\geq 0$, and $\tau_{ch}$ is the charging time.
We assume $Q_{disch}(x,t=0)=Q_{ch}(x,\tau_{ch})$, and $Q_{disch}(x=0,t)=0$
at all times.
The fit parameters are $Q_{0}$ and $D$, although $Q_{0}$ is no longer a
parameter when the data are
normalized.  It is also necessary to allow
the duration of charging $\tau_{ch}$ and the time $t_{1}$ of the
first line scan to vary, as discussed below. For subsequent scans we use
$t_{j}=t_{1}+3(j-1){(min)}$, for j = 1 to 21.

Good agreement between calculated and measured curves is obtained.
Fig.3 shows every third curve $j = 1, 4, \ldots, 19$.
From the best fit $D = (2.8 \pm 0.2)\times 10^{-3}\mu{m}^{2}/s$, $\tau_{ch} \approx 340 min$,
and $t_{1} = 0.35 min$.  The value of D is robust to the variation
of other fit parameters.
The actual measured values $\tau_{ch}= 60 min $ and $t_{1} = 0.05 min$ are much
shorter than the best-fit values.
Allowing $\tau_{ch}$ and $t_{1}$ to be larger than measured has the effect
of broadening the initial charge distribution.  This may be
compensating for our finite spatial resolution, not included in
the model. For $j\geq 2$, the difference in $t_{1}$ is insignificant.  

The same NC film was also imaged during illumination with green laser light (with energy
above the CdSe NC bandgap).  From the fits to the discharging
data in this case $D = (4.9 \pm 0.2)\times 10^{-3}\mu{m}^{2}/s$.

As argued above, the charging data in Fig. 2(a) are not recorded at a constant height.
Specifically, the applied voltage $V_{dc}$ causes the source to appear larger than for $V_{dc} = 0$; 
the apparent increase $\sim 1.4$ times of the electrode height
implies that the height at which the EFM signal is recorded varies from
200nm to 245nm across the scan.  From
the measured z-dependence $\Delta{\phi}(x,y) \propto V_{dc}^{2}/z^{1.3}$,
we infer that the EFM signal above the electrode is $\sim 80 \%$ of the
value expected for a specified height of 200 nm.

Figure 4 shows the EFM images, under the same conditions as for Fig. 2, for
a device with $800 \mu m$ long parallel electrodes, separated by $\sim 4 \mu m$.
The NC diameters are $\approx 4.7nm$ \cite{Nathan}. Fig. 4 (a) shows the EFM image after 
charging of the film for 48 min, while Fig. 4 (b) shows the discharging 
of the film, previously charged for $\approx 2 hrs$.
The four images are slices ($0<y<5 \mu m$) of EFM scans which are
begun 0, 3, 6 and 9 min after $V_{dc}$ is turned off. 
Fig. 4 (c) shows the measured and calculated EFM signal plotted along
the x-axis for $y = 4\mu m$ from the top of the scan, during the
discharging.  Line scans are shown for $t_{1} = 0.33 min,
t_{j}=t_{1}+3(j-1){(min)}$, for $j = 1$ to $4$. The calculated
curves are from a 1D diffusion and we use
$t_{1} = 1.8 min$, larger than the actual value as for Fig. 3.  We 
find $D = 3.6 \times 10^{-3}\mu{m}^{2}/s$ for the film.

The resistance per square of the film $R=1/DC$, where C is the capacitance per unit area.  
The capacitance $C = \epsilon_{0}\epsilon_{r}/h \approx 10^{-16} F/{\mu{m}}^{2}$, where
$\epsilon_{r} \approx 4$ for $SiO_{2}$, and $h=350nm$ is the
thickness of the $SiO_{2}$ layer. For $D = 3 \times 10^{{-3}}{\mu{m}}^{2}/s$,
$R\approx{3} \times {10^{18}}\Omega$ per square.  Even for the geometry
in Fig. 2, this corresponds to $R\sim {10^{16}}\Omega$ and a steady-state current $I\sim 10^{-15} A$ at
$-40V$.  This could not have been detected in previous experiments 
\cite{Morgan}. We estimate that this technique could be used to 
measure resistances per square as high as $\sim {10^{20}}\Omega$ using thicker oxides.  

In conclusion, we have imaged the charge transport
in films of CdSe NCs, and have measured the diffusion coefficient
directly. This method can be applied to systems
with currents so small that they cannot be measured by conventional 
methods. It will also allow quantitative studies of anomalous diffusion, 
which may result from electron interactions.  
Higher spatial and temporal resolution can be achieved using thinner 
oxides and small sample areas, respectively.

We thank D. Novikov and L. Levitov for helpful discussions.
This work was supported by the NSF under MRSEC (DMR 02-13282) and under 
NSEC (PHY-0117795).

{\center{ \section{\bf{Figure Captions}}}}

\section{}

{\bf Figure 1.}  Schematic diagram of the device and the EFM measurement
setup.  CdSe NCs are deposited on top of a device described in the text.
For voltage $V_{dc}<0$ on the source electrode electrons are injected into 
the film from the source; no
charge injection is observed for positive voltages on the source. $q$ is
the charge on the sample, while
$q_{1}$, and $q_2$ are charges induced by $q$ in the gate and in
the tip, respectively.
\section{}
{\bf Figure 2. (COLOR)}  (a) Charging and (b) discharging of a CdSe
film as a function of time as described in the text.
A topographic contour containing the two
electrodes is shown. The circle to the right of the electrodes is a 
flaw in the NC film.
\section{}

{\bf Figure 3.}  The measured and calculated EFM signal
plotted along the x-axis during the discharging
of the film. The calculated curves $(Q(x,t_{j}))^{2}$ are obtained from
the best fit of a diffusion model, described in the text. 
The diffusion coefficient from the fit is $D = (2.8 \pm
0.2)\times 10^{-3}\mu{m}^{2}/s$.  The comparison between data and theory 
shown for $t_{j}$, $j = 1, 4, \dots,19$. Inset: Positions of maxima 
$x_{max,j}$ vs time $t_{j}$, $j=1$ to $21$, and a linear 
fit $x_{max}^{2} = 4Dt$, giving the same value of $D$.

\section{}

{\bf Figure 4. (COLOR)}  EFM images for a  device with long parallel electrodes. (a)
Charging of the film, 48 min after $V_{dc}= -40 V$ is turned on. The
topographic contours of the electrodes are shown. (b) Slices of EFM 
images ($0<y<5 \mu m$) showing discharging of the film 0, 3, 6 and 9 min after $V_{dc}$ is turned off. 
(c) Measured and calculated
EFM signal plotted along the x-axis as described in the text.

\newpage
\center
\section{}

\end{document}